\documentclass[fleqn,12pt,a4paper]{article}

\usepackage{epsfig,amsmath,amssymb,bm}
\usepackage{color}

\begin{document}

%------------------------------------------------------------------

\title{\bfseries Linearized Torsion Waves in a Tensor-Tensor Theory of Gravity}
\author{ \bfseries Chih-Hung Wang  \thanks{Department of Physics,
National Central University, Chungli 320, Taiwan, ROC  (email : chwang@mail.phy.ncu.edu.tw)}}

\maketitle

\def\IR{\mathbb{R}}
\def\be{\begin{equation}}
\def\ee{\end{equation}}
\def\bea{\begin{eqnarray}}
\def\eea{\end{eqnarray}}
\def\ba{\begin{array}}
\def\ea{\end{array}}
\def\bem{\begin{displaymath}}
\def\enm{\end{displaymath}}
\def\d{\textrm{d}}
\def\c{\chi}
\def\va{\varepsilon}
\def\e{\textrm{exp}}
\def\k{\textrm{k}}
\def\C{\textrm{C}}
\def\T{\textrm{T}}
\def\B{\textrm{B}}
\def\P{\textrm{P}}
\def\R{\textrm{R}}
\def\D{\textrm{D}}
\def\i{\textrm{i}}
\def\r{\textrm{r}}
\def\w{\wedge}
\def\lx{\mathcal{L}_X}
\def\AA{\mathcal{A}}
\def\BB{\mathcal{B}}
\def\CC{\mathcal{C}}
\def\DD{\mathcal{D}}
\def\wt{\bar}
\def\h{\hat}
\def\A{\textrm{A}}
\def\up{\stackrel}
\def\1{\frac{1}{2}}
\def\pa{\partial}
\def\q{\textbf{q}}
\def\t{\hat}
\def\w{\wedge}
\def\wi{\tilde}
\def\dd{\dot}
\def\h{\bar}
\def\nn{\nonumber}
\def\S{\mathbf{S}}

\begin{abstract}
 
We investigate a linearized tensor-tensor theory of gravity with torsion and a perturbed torsion wave solution is discovered in background Minkowski spacetime with zero torsion. Furthermore, gauge transformations of any perturbed tensor field are derived in general background non-Riemannian geometries. By calculating autoparallel deviations,  both longitudinal and transverse polarizations of the torsion wave are discovered.
 
\end{abstract}

\section{Introduction}

It is well known that
gravitational energy cannot be localized in general relativity (GR) according to the
equivalence principle. Moreover, the conventional expressions for
gravitational energy-momentum density are non-covariant and vanish
locally in Riemann normal coordinates \cite{MTW} \cite{LL75}. By
exploring the mathematical analogy of gravitational and
electromagnetic fields, the Bel tensor $\h{\B}$ is defined by
\bea
\h\B= \star\, \h{\tau}_{bcd} \otimes e^b \otimes e^c \otimes e^d
\label{Bel-Robinson2}
\eea with associated 3-forms
\bea
\h{\tau}_{bcd}=\frac{1}{2}\left((\i_{X_b}\h{\R}_{cq} \w \star\,
\h{\R}_d{^q} - \h{\R}_{cq} \w \i_{X_b} \star \h{\R}_d{^q}) +
(\i_{X_b}\h{\R}_{dq} \w \star\, \h{\R}_c{^q} - \h{\R}_{dq} \w
\i_{X_b} \star \h{\R}_c{^q}) \right). \nn
\eea  $\{e^a\}$ is an orthonormal co-frame with its dual $\{X_a\}$ and $\h{\R}_{ab}$ are Riemann curvature 2-forms.\footnote{Greek indices $a, b, \cdots$ run over 0 to 3 and Latin indices $\alpha, \beta, \cdots$ over 1 to 3.} $\star$ denotes Hodge map and $\i_{X}$ the interior derivative. $\h{\B}$ has a number of properties in common with the stress-energy tensor of electromagnetic fields \cite{Bel58} \cite{Robinson97}
\cite{BS97} \cite{Garecki01}. By using differential forms, it can easily be proved that $\h{\B}$ is totally symmetric, traceless, and $\h{\nabla} \cdot \h{\B}=0$ in Ricci-flat spacetime \cite{wang06}. $\h{\nabla}$ is Levi-Civita connection. Since Bel tensor holds these interesting properties, it has long been considered as gravitational energy density. Dereli and Tucker (DT) \cite{DT04} proposed a tensor-tensor theory of gravity by introducing a symmetric tensor field $\Phi$ in non-Riemannian spacetime with metric-compatible connection. In DT theory,  generalized Bel 3-forms ${\tau}_{abc}$\footnote{${\tau}_{abc}$ is defined in terms of full curvature 2-forms $\R_{ab}$. In the torsion-free metric-compatible connection, ${\tau}_{abc}$= $\h{\tau}_{abc}$} are naturally appeared in field equations given by 
\bea
- \frac{1}{2\kappa}\,\R_{ab} \w \star e^{ab}{_c}&=&
\Phi^{ab}\tau_{cab} + \lambda \tau^{\Phi}_c, \label{DT-FE1}\\
\frac{1}{2\kappa}\,T^c \w \star \,e_{abc} &=& \lambda(\,\Phi_{bc}
\star \D\Phi^c{_a} - \Phi_{ac} \star \D\Phi^c{_b}\,) \nn\\
&&+\,\, \1\,\{\,\D (\Phi_{ac}\star \R^c{_b}) -
\D(\Phi_{bc}\star\R^c{_a})\}, \label{DT-FE2}\\
\lambda \D\star\D\Phi_{ab} &=& \1\,\R_{ac} \w \star \R_b{^c}
\label{DT-FE3}
\eea where 
\bea
\tau^{\Phi}_c&=&\1 (\i_{X_c}\D \Phi^{ab}\,\star \D \Phi_{ab} + \D
\Phi_{ab} \w \i_{X_c}\star \D \Phi^{ab}), \nn
\eea and $e^{a\ldots b}{_{c\ldots d}}$
denotes $e^a \w \ldots \w e^b \w e_c \w \ldots \w e_d$. $\kappa$ and $\lambda$ are constants, $\Phi_{ab}$ components of $\Phi$ with respect to $\{e^a\}$, $T^a$ torsion 2-forms and $\D$ covariant exterior derivative \cite{BT87}.

In GR,  gravitational waves were first predicted from a linearized theory. Their polarizations have been fully investigated \cite{MTW}. By applying perturbation analysis in DT theory, we also obtain a wave-like solution, i.e. torsion waves and $\Phi$ field waves.  The polarizations of torsion waves turn out to be quite different from the gravitational waves in GR. In section 2, we derive linearized field equations in background curvature and torsion vanishing. These linearized equations are expressed in terms of coordinate-free language. In section 3, we apply Minkowski coordinates on these equations and then obtain a torsion wave and a $\Phi$ wave solution. In section 4, we derive gauge transformations of perturbed field variables in non-Riemannian geometries. Using gauge transformations, some components of torsion waves can be eliminated. In section 5, we calculating the autoparallel deviation in the torsion wave spacetime and discuss polarizations of the torsion wave.

%%%%%%%%%%%%%%%%%%%%%%%%%%%%%%%%%%%%%%%%%%%

\section{Linearized Field Equations}

In our perturbation scheme, we expand field variables $e^a$, $\omega^a{_b}$ and $\Phi_{ab}$ with respect to a dimensionless small parameter $\varepsilon$. When
$\varepsilon=0$, they return to the following background field 
\bea
\wi{\R}_{ab} &=& \d \wi{\omega}_{ab} + \wi{\omega}_{ac} \w
\wi{\omega}^c{_b}=0,\\
\wi{T}^{a} &=& \d \wi{e}^{a} + \wi{\omega}^a{_c} \w \wi{e}^c= 0
\eea with a solution $\wi{\Phi}_{ab}$ satisfying
\bea
\wi{\D}\,\wi{\Phi}_{ab}=0
\eea
where
\bea
\wi{\Phi}_{ab}&=& \wi{\Phi}( \wi{X}_a, \, \wi{X}_b ).\nn
\eea $\wi{e}^a$ and $\wi{\omega}^a{_b}$ denote the background
$\wi{g}$-orthonormal co-frame and connection 1-forms, and
$\wi{X}_a$ is a dual of $\wi{e}^a$. $\wi{\D}$ and $\wi{\star}$ are
covariant exterior derivative and the Hodge map associated to
$\wi{\omega}^a{_b}$ and $\wi{e}^a$, respectively. This background field which represent a flat and torsion-free spacetime with covariant constant $\wi{\Phi}$ is an exact solution of field equations.  We substitute field variables 
\bea
e^a &=& \wi{e}^a + \varepsilon\,\dot{e}^a + \ldots, \nn\\
\omega^a{_b} &=&  \wi{\omega}^a{_b} +
\varepsilon\,\dot{\omega}^a{_b} + \ldots,
\nn\\
\Phi_{ab} &=& \wi{\Phi}_{ab} + \varepsilon\,\dot{\Phi}_{ab} + \ldots \nn
\eea into field equations (\ref{DT-FE1})-(\ref{DT-FE3}) to obtain linearized field equations 
\bea
\wi{\D}\,\dd{\h{\omega}}^{ab} \w \wi{\star}\,\wi{e}_{abc} &=&
-\wi{\D}\,\dd{K}^{ab} \w \wi{\star}\,\wi{e}_{abc}, \label{LFE1}\\
\frac{1}{2\kappa}\,\dd{T}^c \w \wi{\star}\,\wi{e}_{abc} &=&
\lambda(\,\wi{\Phi}_{bc}\,
\wi{\star}\, \up{\cdot}{[\D\Phi^c{_a}]} - \wi{\Phi}_{ac} \,\wi{\star} \up{\cdot}{[\D\Phi^c{_b}]}\,), \nn\\
&&+\,\, \1\,(\,\wi{\Phi}_{ac}\,\wi{\D} \,\wi{\star}
\,\wi{\D}\,\dd{\omega}^c{_b} -
\,\wi{\Phi}_{bc}\,\wi{\D}\,\wi{\star}\,\wi{\D}\,\dd{\omega}^c{_a})\label{LFE2}\\
\wi{\D}\,\wi{\star}\,\wi{\D}\,\dd{\Phi}_{ab} &=& \wi{\Phi}_{cb}\,
\wi{\D}\,\wi{\star} \,\dd{{\omega}}^c{_a} +  \wi{\Phi}_{ca}\,
\wi{\D}\,\wi{\star} \,\dd{{\omega}}^c{_b} \label{LFE3}
\eea where
\bea
\up{\cdot}{[\D\Phi_{ab}]} = \wi{\D} \dd\Phi_{ab} -
\dd\omega^c{_a}\,\wi{\Phi}_{cb} - \dd\omega^c{_b}\,\wi{\Phi}_{ac}.
\eea  $\dd{\h{\omega}}^{ab}$ are perturbed torsion-free connection 1-forms and $\dd{K}^{ab} $ perturbed contorison 1-forms. (\ref{LFE1}) is considered as linearized Einstein equations with source terms $\wi{\D}\,\dd{K}^{ab} \w \wi{\star}\,\wi{e}_{abc}$. It is interesting to notice that
$\wi{\D}\,\wi{\star}\,\wi{\D}\,\dd{\Phi}_{ab}$ are associated with the covariant d'Alembertian operator $\wi{\nabla}\cdot\wi{\nabla}$
defined by
\bea
(\wi{\nabla}\cdot\wi{\nabla}\, \T )(X_1,\ldots, X_r,\, e^1, \ldots,
e^s) \equiv \wi{\nabla}_{X_a}(\wi{\nabla}\,\T)\, ( X^a, X_1,\ldots,
X_r,\, e^1, \ldots, e^s) \nn
\eea where $\T \in \Gamma T^s_r M$ is a tensor field. Using the
definition of ${\nabla}\,\T$
\bea
(\nabla\,\T)\,(X,\,X_1,\ldots, X_r,\, e^1, \ldots, e^s )
=(\nabla_X \T)\,(X_1,\ldots, X_r,\, e^1, \ldots, e^s ) \nn
\eea and
\bea
\D S_{a\ldots b}{^{c\ldots d}}= \{(\nabla_{X_j} S)\, (X_a,\ldots,
X_b, e^c, \ldots, e^d)\}\,e^j
\eea for arbitrary 0-forms $S_{a\ldots b}{^{c\ldots d}}$, we
obtain
\bea
\wi{\D}\,\dd{\Phi}_{ab} = \{(\wi{\nabla} \Phi_P)\, (
\wi{X}_j,\,\wi{X}_a, \wi{X}_b)\} \,e^j \label{DstarD}
\eea where
\bea
\Phi_P = \dd{\Phi}_{ab}\,\wi{e}^a \otimes \,\wi{e}^b
\eea and  $\wi{\nabla}$ denotes the background connection. By performing $\wi{\D}\,\wi{\star}$ on (\ref{DstarD}) gives
\bea
\wi{\D}\,\wi{\star}\,\wi{\D}\,\dd{\Phi}_{ab} &=&
\wi{\D}\left((\wi{\nabla} \Phi_P)\, (\wi{X}_j,\,\wi{X}_a,
\wi{X}_b)\right) \w \wi{\star}\,\wi{e}^j \nn\\
&=&\wi{\nabla}_{\wi{X}_c} \,(\wi{\nabla}\Phi_P)\,
(\wi{X}^c,\,\wi{X}_a, \wi{X}_b)\, \wi{\star}\, 1 \nn\\
&=& (\wi{\nabla}\cdot \wi{\nabla}\Phi_P)\,(\wi{X}_a, \wi{X}_b).
\eea  
So (\ref{LFE3}) is a non-vacuum
covariant wave equation.  In the following section, we find a torsion wave solution using the ansatz  $\dd{\omega}_{ab}=0$.

%%%%%%%%%%%%%%%%%%%%%%%%%%%%%%%%%%%%%%%%%%

\section{A Torsion-Wave Solution}

In previous section, we chose the background field $\wi{\R}_{ab}=\wi{T}^a=0$ with the background tensor field $\wi{\Phi}$
satisfying $\wi{\D}\wi{\Phi}_{ab}=0$. Because $\wi{\Phi}_{ab}$
depends on $\{\wi{e}^a\}$, it simplifies our calculations to
choose the background orthonormal co-frame as $\wi{e}^a= \d x^a$,
i.e. Minkowski metric in Minkowski coordinates $\{x^a= t, x^\alpha\}$. So
$\wi{\omega}_{ab}=0$ and $\wi{\D}$ becomes the exterior derivative
$\d$. According to the equation $\d \wi{\Phi}_{ab}=0$, we obtain
that components $\wi{\Phi}_{ab}$  should be constant
in this background orthonormal co-frame $\{\d x^a\}$. We then have
the following background fields
\bea
\wi{\R}_{ab}&=&0, \nn\\
\wi{T}^a&=&0, \nn\\
\wi{e}^a&=& \d x^a \label{backgroundfield0}
\eea and choose
\bea
\wi{\Phi}_{ab} &=&
\textrm{diag}(\Phi_0,\,\Phi_1,\,\Phi_1,\,\Phi_1)
\label{backgroundfield}
\eea where $\Phi_0$ and $\Phi_1$ are constants. Furthermore, Using the ansatz
\bea
\dd T^a = \d \dd e^a, \label{ansatz-eq}
\eea i.e. $\dd{\omega}^a{_b}=0$, (\ref{LFE1})-(\ref{LFE3}) are decoupled and then reduce to
\bea
\frac{1}{2\kappa}\,\d\dd{e}^c \w \wi{\star}\,\wi{e}_{abc} &=&
\lambda(\,\wi{\Phi}_{bc}\, \wi{\star}\,\d\dd{\Phi}^c{_a} -
\,\wi{\Phi}_{ac}\, \wi{\star}\,\d\dd{\Phi}^c{_b}), \label{1-stconnection-d}\\
\d\,\wi{\star}\,\d\dd{\Phi}_{ab}&=&0. \label{1-sttensorfield-d}
\eea We can first solve (\ref{1-sttensorfield-d}) and then
substitute the solution $\dd{\Phi}_{ab}$ into
(\ref{1-stconnection-d}) to obtain a solution of $\dd{e}^a$ and
$\dd{T}^a$. By solving the vacuum symmetric tensor wave equation
(\ref{1-sttensorfield-d}), a particular solution, monochromatic
plane-wave solution, are obtained
\bea
\dd{\Phi}_{ab}= \Re[A_{ab} \,\e(i\,k_p x^p)] \label{solutionPhi}
\eea with the amplitude $\A=A_{ab} \d x^a \otimes \d x^b$ and the
wave vector $\k= k^a \pa_a$ satisfying
\bea
A_{ab}&=& A_{ba}, \hspace{1.7cm} (\textrm{A is symmetric})\\
k_p k^p&=&0,  \hspace{2.0cm} (\textrm{k is a null vector})
\eea where $\Re$ denotes the real part of the expression in
brackets and $A_{ab}$, $k_p$ are constants. $\{\pa_a\}$ is the
dual of $\{\d x^a\}$. 

In order to solve (\ref{1-stconnection-d}),
we write down these equations explicitly
\be\ba{lll}
-(\i_{\pa_0}\i_{\pa_1}\d \dd{e}^c )\,\wi{\star}\, \d x_c +
(\i_{\pa_0}\i_{\pa_c}\d \dd{e}^c )\,\wi{\star}\, \d x_1 -
(\i_{\pa_1}\i_{\pa_c}\d \dd{e}^c )\,\wi{\star}\, \d x_0 &=&
\alpha \,\wi{\star}\,\d\dd{\Phi}_{01},\\
-(\i_{\pa_0}\i_{\pa_2}\d\dd{e}^c )\,\wi{\star}\, \d x_c +
(\i_{\pa_0}\i_{\pa_c}\d \dd{e}^c )\,\wi{\star}\, \d x_2 -
(\i_{\pa_2}\i_{\pa_c}\d \dd{e}^c )\,\wi{\star}\, \d x_0 &=&
\alpha\,\wi{\star}\,\d\dd{\Phi}_{02}, \\
-(\i_{\pa_0}\i_{\pa_3}\d \dd{e}^c )\,\wi{\star}\, \d x_c +
(\i_{\pa_0}\i_{\pa_c}\d \dd{e}^c )\,\wi{\star}\, \d x_3 -
(\i_{\pa_3}\i_{\pa_c}\d \dd{e}^c )\,\wi{\star}\, \d x_0 &=& \alpha
\,\wi{\star}\,\d\dd{\Phi}_{03}, \label{1-stconnection-d1-1}
\ea\ee
\be\ba{lll}
-(\i_{\pa_1}\i_{\pa_2}\d \dd{e}^c )\,\wi{\star}\, \d x_c +
(\i_{\pa_1}\i_{\pa_c}\d \dd{e}^c )\,\wi{\star}\, \d x_2 -
(\i_{\pa_2}\i_{\pa_c}\d \dd{e}^c )\,\wi{\star}\, \d x_1 &=& 0, \\
-(\i_{\pa_1}\i_{\pa_3}\d \dd{e}^c )\,\wi{\star}\, \d x_c +
(\i_{\pa_1}\i_{\pa_c}\d \dd{e}^c )\,\wi{\star}\, \d x_3 -
(\i_{\pa_3}\i_{\pa_c}\d \dd{e}^c )\,\wi{\star}\, \d x_1 &=& 0, \\
-(\i_{\pa_2}\i_{\pa_3}\d \dd{e}^c )\,\wi{\star}\, \d x_c +
(\i_{\pa_2}\i_{\pa_c}\d \dd{e}^c )\,\wi{\star}\, \d x_3 -
(\i_{\pa_3}\i_{\pa_c}\d \dd{e}^c )\,\wi{\star}\, \d x_2 &=& 0
\label{1-stconnection-d1-2}
\ea\ee where
\bea
\alpha=2\,\kappa\lambda(\Phi_0+\Phi_1), \nn
\eea and $\d x_a = \eta_{ab}\,\d x^b$. The equations
(\ref{1-stconnection-d1-2}) can be considered as constraint
equations for $\d \dd{e}^a$. In general,
\be\ba{lll}
\d \dd{e}^0 &=& \AA_{ab}\, \d x^a \w \d x^b, \\
\d \dd{e}^1 &=& \BB_{ab}\, \d x^a \w \d x^b, \\
\d \dd{e}^2 &=& \CC_{ab}\, \d x^a \w \d x^b, \\
\d \dd{e}^3 &=& \DD_{ab}\, \d x^a \w \d x^b \label{dea}
\ea\ee where $\mathcal{A}_{ab}$, $\mathcal{B}_{ab}$, $
\mathcal{C}_{ab}$ and $ \mathcal{D}_{ab}$ are functions of $x^a$. Substituting (\ref{dea}) into (\ref{1-stconnection-d1-2}) gives the following conditions 
\bea
\AA_{12}=\AA_{13}=\AA_{23}=\BB_{23} = \CC_{13} = \DD_{12}=0,\nn\\
\AA_{01}=\CC_{12}=\DD_{13}, \,\,\,\AA_{02}
=-\BB_{12}=\DD_{23},\,\,\, \AA_{03}= - \BB_{13} =
-\CC_{23}\label{dea2}
\eea and the 24 unknown functions reduce to 12. For solving
(\ref{1-stconnection-d1-1}), we consider the plane-wave of $\Phi$ propagating along the $x^1$
direction and then (\ref{solutionPhi}) becomes
\bea
\dd{\Phi}_{ab}= \Re[A_{ab} \,\e \{- i\,k( x^0 - x^1)\}]. \label{Phi_x}
\eea Substituting (\ref{Phi_x}) into (\ref{1-stconnection-d1-1}) gives a particular solution
\bea
\dd{e}^0&=&- \1\,\alpha\,\dd{\Phi}_{01}\, \d x^0, \nn\\
\dd{e}^1&=& \left(\mathcal{P}(x^0) - \alpha\,\dd{\Phi}_{01} \right)\,\d x^0\,+\, \1\,\alpha\,\dd{\Phi}_{01}\, \d x^1, \nn\\ 
\dd{e}^2&=& \CC(x^0)\,\d x^0\,+\, \1\,\alpha\,\dd{\Phi}_{01}\, \d x^2,\nn\\
\dd{e}^3&=& \DD(x^0)\,\d x^0\,+\, \1\,\alpha\,\dd{\Phi}_{01}\, \d x^3  \label{dea3}
\eea with an extra condition on $\dd{\Phi}_{ab}$
\bea
A_{02}=A_{03}=0
\eea where $\mathcal{P}(x^0)$, $\CC(x^0)$, and $\DD(x^0)$ are
still unknown functions. We can also obtain the perturbed torsion
wave solution
\bea
&&\dd{T}^0 = \d \dd{e}^0 = \1\,\alpha\,\Re[A_{01} \,f_1]\,\d x^0 \w \d x^1, \nn\\
&&\dd{T}^1= \d \dd{e}^1 = \1\,\alpha\,\Re[A_{01} \,f_1]\,\d x^0 \w \d x^1, \nn\\
&&\dd{T}^2= \d \dd{e}^2 = \1\,\alpha\,\Re[A_{01} \,f_1]\,(- \d x^0 \w \d x^2 \,+ \,\d x^1 \w \d x^2),\nn \\
&&\dd{T}^3= \d \dd{e}^3 = \1\,\alpha\,\Re[A_{01} \,f_1]\,(- \d x^0
\w \d x^3 \,+ \,\d x^1 \w \d x^3) \label{torisonwavesolution}
\eea where $f_1 = i\,k\,\e \{- i\,k(x^0 - x^1)\}$

From  (\ref{1-stconnection-d}), we may expect that perturbed
orthonormal co-frames $\dd{e}{^a}$ allow gauge transformations
\bea
\dd{e}{^a} \rightarrow \dd{e}{^a} + \d\,V^a \label{gaugetrano}
\eea where $V^a$ are 0-forms.  These unknown functions
$\mathcal{P}(x^0)$, $\CC(x^0)$, and $\DD(x^0)$ can be transformed
away by performing the gauge transformations (\ref{gaugetrano}).
In addition to gauge transformations of $\dd{e}{^a}$, we have to
know the gauge transformations of $\dd{\Phi}_{ab}$. In the
following section, we explore the gauge transformations of
perturbed field variables $\dd{e^a}$, $\dd{\omega}_{ab}$,
$\dd{T}^a$, $\dd{R}^a{_b}$, $\dd{\Phi}_{ab}$ in general background
fields.

%%%%%%%%%%%%%%%%%%%%%%%%

\section{Gauge Transformations}

The formulation of perturbations of spacetimes is developed in
\cite{SW74} which is based on the work in \cite{geroch69}. We
adopt their perturbation formulation to discover the gauge
transformations of  $\dd{e^a}$,
$\dd{\omega}_{ab}$, $\dd{T}^a$, $\dd{R}^a{_b}$, $\dd{\Phi}_{ab}$
in non-Riemannian spacetime. In \cite{SW74} , a gauge
transformation of any perturbed tensor field $\dd{\T}$ can be defined in terms of Lie derivative $\mathcal{L}_X$
\bea
\dd{\T}^\prime = \dd{\T} + \, \mathcal{L}_X \wi{\T},
\eea where $X$ is any vector field on spacetime manifold $M$ and
$\wi{\T}$ denotes the background tensor field. If $\mathcal{L}_X \wi{\T}$ vanishes for all $X \in M$, we say that $\dd{\T}$ is gauge
invariant. We first calculate the gauge
transformations of $\dd{e}^a$. By expanding metric $g$
with respect to $\varepsilon$ gives
\bea
g = \wi{g} \,+ \varepsilon \,\dd{g} \,+\, O(\varepsilon^2)
\eea with
\be\ba{lll}
\wi{g}&=& \wi{e}^a \otimes \wi{e}_a, \nn\\
\dd{g}&=& \dd{e}^a \otimes \wi{e}_a + \wi{e}_a \otimes \dd{e}^a
\nn
\ea\ee and using
\bea
\mathcal{L}_V \wi{e}^a &=& \wi{\D} V^a + \i_V \wi{T}^a + \wi{\nabla}_V \wi{e}^a
\eea where $\wi{\nabla}_V$ denotes the background connection, we
obtain \footnote{$\wi{\nabla}_V\wi{g}$ vanishes because of metric-compatible connection $\wi{\nabla}$}
\bea
\mathcal{L}_V \wi{g} &=& (\wi{\D} V^a + \i_V \wi{T}^a) \otimes \wi{e}_a + \wi{e}_a
\otimes (\wi{\D} V^a + \i_V \wi{T}^a)
\eea  where $V$ is a vector field on $M$. So gauge transformations of $\dd{e}^a$ are
\bea
\dd{e}^a{^\prime} = \dd{e}^a + \wi{\D} V^a + \i_V \wi{T}^a.
\label{gaugetransformatione}
\eea 

 Expand $(3,0)$ torsion tensor $T$ with respect to $\varepsilon$
\bea
T= 2\,\wi{T}_a \otimes \wi{e}^a \,+ \,2\,\varepsilon (\dd{T}_a \otimes
\wi{e}^a + \wi{T}_a \otimes \dd{e}^a) + O(\varepsilon^2)
\eea and then calculate $\mathcal{L}_V \wi{T}$ to obtain
\bea
\mathcal{L}_V \wi{T} &=& 2\,\{(\wi{\D} \i_V \wi{T}^a + \i_V \wi{\R}{^a}{_c} \w \wi{e}^c
+ V^c\wi{\R}{^a}{_c}) \otimes \wi{e}_a + \wi{T}_a \otimes (\wi{\D}
V^a + \i_V \wi{T}^a)\}. \nn
\eea So gauge transformations of $\dd{T}^a$ are
\bea
\dd{T}^a{^\prime}= \dd{T}^a + \wi{\D} \i_V \wi{T}^a + \i_V
\wi{\R}{^a}{_c} \w \wi{e}^c + V^c\wi{\R}{^a}{_c}.
\label{gaugetransformationtorsion}
\eea It is easily to show that perturbed connection 1-forms are gauge invariant, i.e. $\dd{\omega}^\prime_{ab} = \dd{\omega}_{ab}$.  We next expand $(4,0)$ curvature
tensor $\R$ with respect to $\varepsilon$ which gives
\bea
\R=2\,\wi{\R}_{ab} \otimes \wi{e}^{ab} + 2\,\varepsilon (\,\dd{\R}_{ab}
\otimes \wi{e}^{ab} + \wi{\R}_{ab} \otimes \dd{e}^a \w \wi{e}^b +
\wi{\R}_{ab} \otimes \wi{e}^a \w \dd{e}^b ) + O(\varepsilon^2) \nn
\eea and using $\mathcal{L}_V \wi{\R}$ we obtain 
\bea
\dd{\R}^{\prime}_{ab} = \dd{\R}_{ab} + \wi{\D} \i_V \wi{\R}_{ab}.
\eea Finally, gauge transformations of $\Phi_{ab}$ are
\bea
\dd{\Phi}^\prime_{ab}= \dd{\Phi}_{ab} + \i_V \wi{\D}
\wi{\Phi}_{ab}.
\eea We have shown gauge transformations of $\dd{e}^a$,
$\dd{T}^a$, $\dd{\omega}_{ab}$, $\dd{\R}{^a}{_b}$, and
$\dd{\Phi}_{ab}$ in general background fields. By considering the
background fields (\ref{backgroundfield0}) and
(\ref{backgroundfield}), we can see that $\dd{T}^a$,
$\dd{\omega}_{ab}$, $\dd{\R}{^a}{_b}$, and $\dd{\Phi}_{ab}$ are
gauge invariant. $\dd{e}^a$ is still a gauge dependent variable
and its gauge transformations are
\bea
\dd{e}^a{^\prime} = \dd{e}^a + \d V^a
\eea which are the same as (\ref{gaugetrano}).

We perform the following gauge transformations
\bea
&&V^0=  \1\,\alpha\,\Re[\,\frac{i}{k}\,A_{01}\,\e \{- i\,k( x^0 -
x^1)\}],\nn\\
&&V^1=  \1\,\alpha\,\Re[\,\frac{i}{k}\,A_{01}\,\e \{- i\,k( x^0 -
x^1)\}] + \  \int_{a_0}^{x^0} ( \,\mathcal{P}(x^0) \d x^0\, ) + \mathcal{P}(a_0)\,x^0,\nn\\
&&V^2= \int_{b_0}^{x^0} \left(\, \CC(x^0) \d x^0\, \right) + \CC(b_0)\,x^0, \nn\\
&&V^3= \int_{d_0}^{x^0} \left(\, \DD(x^0) \d x^0\, \right) + \DD(d_0)\, x^0
\eea where $a_0$, $b_0$, $d_0$ are constants, so 
(\ref{dea3}) becomes
\bea
&&\dd{e}^0{^\prime} = - \1\,\alpha\,\Re[A_{01} \,\e \{- i\,k( x^0
- x^1)\}]\d
x^1=  - \1\,\alpha\,\dd{\Phi}_{01}\d
x^1, \nn\\
&&\dd{e}^1{^\prime}= -\1\, \alpha\,\Re[A_{01} \,\e \{- i\,k( x^0 -
x^1)\}]
\d x^0 = - \1\,\alpha\,\dd{\Phi}_{01}\d
x^0, \nn\\
&&\dd{e}^2{^\prime}=  \1\,\alpha\,\Re[A_{01} \,\e \{- i\,k( x^0 -
x^1)\}]\,
\d x^2=\1\,\alpha\,\dd{\Phi}_{01}\d
x^2, \nn\\
&&\dd{e}^3{^\prime}=  \1\,\alpha\,\Re[A_{01} \,\e \{- i\,k( x^0 -
x^1)\}]\, \d x^3=\1\,\alpha\,\dd{\Phi}_{01}\d
x^3 \label{ddea}
\eea where $\CC(x^0)$, $\DD(x^0)$ and $\mathcal{P}(x^0)$ have been
transformed away. 

We compare (\ref{ddea}) to the linearized gravitational wave in GR. In the linearized GR, two different polarizations of the
plane-wave solutions in transverse and traceless gauges have been
found \cite{MTW} and by considering the plane-wave propagating along $x^1$,
the solution gives
\bea
\dd{g}_{GR}&=& h^1_{GR} \,\d x^2 \otimes \d x^2 -  h^1_{GR} \,\d
x^3 \otimes \d x^3, \nn\\
\dd{g}_{GR}&=& h^2_{GR} \,\d x^2 \otimes \d x^3 +  h^2_{GR} \,\d
x^3 \otimes \d x^2 \label{perturbedmetricGR}
\eea with
\bea
h^1_{GR}=\Re[A_{+} \,\e \{- i\,k( x^0 - x^1)\}] \nn\\
h^2_{GR}=\Re[A_{\times} \,\e \{- i\,k( x^0 - x^1)\}]
\eea where $A_{+}$ and $A_{\times}$ are amplitudes. The torsion wave solution  is  
\bea
\dd{g}^\prime &=& h\,\d x^2 \otimes \d x^2 + h\, \d x^3 \otimes \d x^3,
\label{perturbedmetric}
\eea where
\bea
h= \alpha\,\Re[A_{01} \,\e \{- i\,k( x^0 - x^1)\}].
\eea If we consider particles following geodesics, one may expect that the polarizations of torsion waves are transverse but not traceless. So we obtain another possible polarization mode of gravitational waves. On the other hands, if we assume particles following autoparallels instead of geodesics, autoparallel deviation should give us different polarizations of torsion waves.  In the next section, we will investigate this possibility.

%%%%%%%%%%%%%%%%%%%%%%%%%%%%%%%

\section{Autoparallel Deviation in a Torsion-Wave Spacetime}

 Consider a family of autoparallels
$\gamma: [0,1] \times [0,1]\rightarrow M$, $\tau,\, s \mapsto
p=\gamma(\tau, s)$ in a spacetime manifold $M$ and tangent vectors $\gamma_*
\partial_\tau$ at every point of $\gamma(\tau,
s)$ satisfying 
\bea
\nabla_{\gamma_*\partial_\tau}\gamma_*\partial_\tau=0
\label{autoparallel}
\eea with affine parametrization
\bea
g(\gamma_*\partial_\tau, \gamma_*\partial_\tau)=-1.
\eea 
We may consider $\gamma_*\partial_\tau|_{\gamma_0}\equiv \dd{C}$ as particle's
4-velocity along $\gamma_0\equiv \gamma(\tau, 0)$ and $\gamma_*\partial_s|_{\gamma_0} \equiv S$ the
separation vector from the particle to its neighbor particles.  
The definition of the $(3,1)$ curvature tensor $\R$ gives
\bea
\nabla_{\dd{C}}\nabla_{S} \dd{C} - \nabla_{S}\nabla_{\dd{C}}
\dd{C} - \nabla_{[\dd{C}, S]} \dd{C} &=& \R(\dd{C}, S, \dd{C},
-)|_{\gamma_0}\nn \\ &\equiv& \t{\R}(\dd{C}, S, \dd{C}, -) 
\label{autoparallel_deviation_1}
\eea where hats indicate evaluation over the image of $\gamma_0$. Using (\ref{autoparallel}), 
\bea
\mathcal{L}_{\gamma_*
\partial_\tau}\gamma_*\partial_s =
\gamma_{*}[\partial_\tau,\,
\partial_s]=0 \label{commutative relation}
\eea 
and the definition of $(2,1)$ torsion tensor
\bea
\nabla_{\dd{C}} S - \nabla_S \dd{C} - [\dd{C}, \,S] =
\t{\T}(\dd{C}, S, -),
\eea (\ref{autoparallel_deviation_1}) becomes
\bea
\nabla_{\dd{C}}\nabla_{\dd{C}}S -
\nabla_{\dd{C}}\left(\t{\T}(\dd{C}, S, -)\right) = \t{\R}(\dd{C},
S, \dd{C}, -) \label{autoparallel_deviation_2}
\eea which is the equation of autoparallel deviation.
$S$ indicates how neighboring particles behave when a torsion wave passes through
them.

In torsion-wave spacetime, we construct
an orthonormal frame $\{X_a\}$
\bea
X_0&=& \partial_0 + \varepsilon\,f \partial_1 + O(\varepsilon^2),\nn\\
X_1&=& \partial_1 + \varepsilon\,f \partial_0 + O(\varepsilon^2), \nn\\
X_A&=& \partial_A - \varepsilon\,f \partial_2 + O(\varepsilon^2) 
\label{X_a1}
\eea 
where $A=2,3$ and
\bea
f=\1\,h = \1\,\alpha\,\Re[A_{01} \,\e \{- i\,k( x^0 - x^1)\}].
\eea Since
\bea
\nabla_{\t{X}_0}\t{X}_0=\t{\omega}^c{_0}(\t{X}_0)\t{X}_c =
O(\varepsilon^2),
\eea it can be identified with the particle's 4-velocity $\dd{C}$
up to first-order in $\varepsilon$, i.e.
\bea
\t{X}_0= \t{\partial}_0 + \varepsilon\,\t{f} \t{\partial}_1 +
O(\varepsilon^2) = \dd{C} + O(\varepsilon^2). \label{ddC}
\eea
Calculating (\ref{autoparallel_deviation_2}) with respect to $\{\t{X}_a\}$ to first-order gives
\bea
\frac{\d^2 S^{\h{a}}}{\d t^2} - \,\varepsilon \frac{\d}{\d
t}\left(\t{\dd{T}}^{\h{a}}{_{\h{0}\h{c}}}\,S^{\h{c}}\right)=0
\label{autoparallel_deviation_4}
\eea where $\t{\dd{T}}^{\h{a}}{_{\h{0}\h{c}}}$ are components of $\dd{T}^a$ with respect to $\{X_a\}$ and $t$ is the coordinate time parameter.\footnote{$ t= \tau +  O(\varepsilon^2)$.} $\h{ }$ on indices denotes the components of tensor fields with respect to $\{X_a\}$. So substituting the  torsion wave solution into
(\ref{autoparallel_deviation_4}), we obtain
\bea
\frac{\d^2 S^{\h{0}}}{\d t^2} &=& \va\,\alpha\,\Re[A_{01}
\,\t{f_1}] \frac{\d S^{\h{1}}}{\d t} + 2\,\va\,k^2\,\t{f}
S^{\h{1}} + O(\va^2),\nn\\
\frac{\d^2 S^{\h{1}}}{\d t^2}&=&  \va\,\alpha\,\Re[A_{01}
\,\t{f_1}]\frac{\d S^{\h{1}}}{\d t} + 2\,\va\,k^2\,\t{f}
S^{\h{1}} + O(\va^2),\nn\\
\frac{\d^2 S^{\h{A}}}{\d t^2} &=& - \va\,\alpha\,\Re[A_{01}
\,\t{f_1}] \frac{\d S^{\h{A}}}{\d t} - 2\,\va\,k^2\,\t{f}
S^{\h{A}} + O(\va^2). \label{autoparallel_deviation_5}
\eea  For solving (\ref{autoparallel_deviation_5}), we expand $S^{\h{a}}$ with
respect to $\va$
\bea
S^{\h{a}}= \wi{S}^{\h{a}} + \va\,\dd{S}^{\h{a}} + O(\va^2)
\eea 
and consider the following initial conditions
\bea
S^{\h{a}}(0)= \wi{S}^{\h{a}} \nn\\
\frac{\d S^{\h{a}}}{\d t}(0)=0.
\eea  The solution for zeroth-order and first-order are
\bea
&&\wi{S}^{\h{0}}=0,\nn\\
&&\wi{S}^{\h{\alpha}}=\textrm{constant}
\eea  and 
\bea
\dd{S}^{\h{0}}&=& - 2\,\t{f}\, \wi{S}^{\h{1}},\nn \\
\dd{S}^{\h{1}}&=& - 2\,\t{f}\, \wi{S}^{\h{1}}, \nn\\
\dd{S}^{\h{A}}&=&  2\,\t{f} \,\wi{S}^{\h{A}}
\eea where 
\bea
\t{f}\,(0)=0, \nn\\
\t{f_1}\,(0)=0.
\eea  So the solutions of (\ref{autoparallel_deviation_5}) to
first-order are
\bea
S^{\h{0}}(t)&=& -2\, \va\,\t{f}\, \wi{S}^{\h{1}},\nn\\
S^{\h{1}}(t)&=& \wi{S}^{\h{1}} - 2\,\va\,\t{f}\, \wi{S}^{\h{1}},\nn\\
S^{\h{A}}(t)&=& \wi{S}^{\h{A}} +  2\,\va\,\t{f}\, \wi{S}^{\h{A}}.
\label{autoparallel_deviation_s}
\eea In the linearized GR, it is well-known that a gravitational
wave is transverse. The solution (\ref{autoparallel_deviation_s})
shows that the torsion wave has not only transverse parts but also
longitudinal parts. From this solution, we can clarify the
polarization of the torsion wave. Consider a particle $\textbf{Q}$
sitting at the centre of a sphere and suppose the particle
$\textbf{Q}$ and every particle on the sphere carries a clock. All
the clocks have been synchronized with respect to background
spacetime (i.e. Minkowski spacetime with coordiantes $\{x^a\}$ and
torsion field vanishing) before the torsion wave comes. Moreover,
all the particle are at rest in the background spacetime. Using
the separation vector $S$, we define the distances of $\textbf{Q}$
and neighboring particles sitting in three different spatial
directions $\t{X}_\alpha$ by $\sqrt{g(S_L, S_L)}$,
$\sqrt{g(S_{T_2}, S_{T_2})}$ and $\sqrt{g(S_{T_3}, S_{T_3})}$,
where
\bea
&&S_L= S^{\h{1}} \t{X}_1, \nn\\
&&S_{T_2} =S^{\h{2}} \t{X}_2, \nn\\
&&S_{T_3} =S^{\h{3}} \t{X}_3. \nn
\eea When a torsion wave propagates along $x^1$ passing through
these particles, the distance between these neighboring particles
will change. The changes of the distance can be obtained by
calculating $g(S, S)$ to first-order which give
\bea
\sqrt{g(S_L, S_L)}= |\wi{S}^{\h{1}}|\,(\, 1 - 2\,\va\,\t{f}\,) + O(\va^2), \label{SL}\\
\sqrt{g(S_{T_2}, S_{T_2})}= |\wi{S}^{\h{2}}|\,(\, 1 + 2\,\va\,\t{f}\,)
+ O(\va^2), \label{ST2}
\\
\sqrt{g(S_{T_3}, S_{T_3})}= |\wi{S}^{\h{3}}|\,(\, 1 + 2\,\va\,\t{f}\,)
+ O(\va^2). \label{ST3}
\eea
 If the distances (to first-order) between the particle $\textbf{Q}$ and those
neighboring particles sitting on the transverse plane are expanded, the longitudinal
distances will be contracted. Attempts to detect
gravitational waves are well underway \cite{schutz99}
\cite{danzmann96} \cite{RH00}. It would also be interesting to
detect longitudinal modes of gravitational waves.

\section{Conclusion}

We first discover a plane wave solution in linearized DT theory. The plane wave solution consists of torsion waves and tensor waves. We use gauge transformations to simplified our solution. To investigate the polarizations of torsion waves, we calculate the equations of motion of particles. We assume particles following time-like autoparallels instead of geodesics and then calculating autoparallel deviation. We then obtain both longitudinal and transverse polarization.  

% We also investigated the gauge transformations of
%first-order perturbed field variables in non-Riemannian spacetime
%(metric-compatible connection with torsion). For observing the
%polarizations of the torsion wave solution, we explored the
%autoparallel deviation of massive spinless particles. The
%approximate solution (to first-order) of autoparallel deviation
%yielded the changes of the distances of the particle \textbf{Q}
%and its neighboring particles when the torsion wave pass through
%these particles. The longitudinal modes of the torsion wave were
%discovered. 

Since gravitational wave observations are underway, and if Laser Interferometer 
Gravitational Wave Observatory (LIGO) \cite{BW99} and Laser Interferometer Space Antenna (LISA)
\cite{danzmann96} \cite{RH00} detect longitudinal modes of
gravitational waves, these calculations may be used to explain the
data.

\paragraph{Acknowledgments\\\\} 
I am very grateful to Prof Robin Tucker for his encouragements and instructions during my PhD study. Moreover, I would also like to thank Dr David Burton for helpful discussion.


\begin{thebibliography}{99}
\footnotesize
{

\bibitem{BW99} Barish, C. $\&$ Weiss, R., \textit{Physics Today}
\textbf{52}, 44, (1999).

\bibitem{Bel58}
Bel, L., \textit{Acad. Sci., Paris} \textbf{247}, 1297 (1958).

\bibitem{BS97}
Bonilla, M. A. G. $\&$ Senovilla, J. M. M., \textit{Gen. Rel.
Grav.} \textbf{29}, 91, (1997).

\bibitem{BT87}
Benn, I. M. $\&$ Tucker, R. W. \textit{An Introduction to Spinors
and Geometry with Applications in Physics.} Adam Hilger (IOP
Publishing Ltd), (1987)

\bibitem{danzmann96} Danzmann, K. \textit{et al.}, \textit{Class. and Quant. Gravity}
\textbf{13}, A247, (1996).

\bibitem{DT04} Dereli, T. $\&$ Tucker,
R. W., \textit{Class. and Quant. Gravity} \textbf{21}, 1459-1464,
(2004).

\bibitem{geroch69}
Geroch, R. \textit{Commun. Math. Phys.}  \textbf{13}, 180, (1969).

\bibitem{Garecki01}
Garecki, J. \textit{Ann. Phys.} (Leipzig) \textbf{10}, 911,
(2001).

\bibitem{LL75}  Landau, L. D., $\&$ Lifshitz, E. M., \textit{The Classical Theory of Fields}, (Butterworth, Heinmann), (2000),
4th Revised English Edition.

\bibitem{MTW} Misner, C. W., Thorne, K. S. and Wheeler, J. A., \textit{Gravitation}, (Freeman, New York,
1973).

\bibitem{RH00} Rowan, S. $\&$ Hough, J., \textit{ Living Rev. Relativ.}
\textbf{3}, (2000).

\bibitem{Robinson97}
Robinson, I., \textit{Class. Quant. Grav.} \textbf{14}, A331
(1997).

\bibitem{schutz99} Schutz, B. F., \textit{Class. and Quant. Gravity} \textbf{16}, A131-A156
(1999).

\bibitem{SW74} Stewart, J. M.  $\&$ Walker, M., \textit{Proc. Roy. Soc. Lond.} \textbf{A341}, 49 (1974)

%\bibitem{Benn82}
%Benn, I. M.  \textit{Ann. Inst. H. Poincar\'e}  \textbf{37}, 67, (1982).

%\bibitem{BT87}
%Benn, I. M. $\&$ Tucker, R. W. \textit{An Introduction to Spinors
%and Geometry with Applications in Physics.} Adam Hilger (IOP
%Publishing Ltd), (1987)

%\bibitem{DT82}
%Dereli, T. $\&$ Tucker, R. W., \textit{Phys. Letts.}
%\textbf{B110}, 206, (1982).

%\bibitem{DT02}
%Dereli, T. $\&$ Tucker, R. W., \textit{Mod. Phys. Lett.} A
%\textbf{17}, 421-428, (2002).

%\bibitem{Perlick91} Perlick, V., \textit{Class. Quant. Grav.} \textbf{8}, 1369, (1991)

%\bibitem{Tucker04}
%Tucker, R. W.  \textit{Proc. Roy. Soc. Lond.} A \textbf{460}, 2818
%(2004).

%\bibitem{Tucker05}
%Tucker, R. W.  \textit{Phil. Mag.}  \textbf{85}, 3911
%(2005).

\bibitem{wang06} Wang, C. H., \textit{Ph.D. thesis}, Lancaster, 2006 (unpublished).
}
\end{thebibliography}
\end{document}